\documentclass[aps,a4paper,preprint,superscriptaddress,preprintnumbers,floatfix,nofootinbib,amsmath,amssymb,showkeys]{revtex4-1}
\usepackage[utf8]{inputenc}
\usepackage{graphicx} 
 \usepackage[table]{xcolor}
  \usepackage{xcolor}
\usepackage{amsmath}
\usepackage{amssymb}
\usepackage{multirow} 

\usepackage{url}
\usepackage{hyperref}
\usepackage{color}
\usepackage{cancel}
\usepackage{cleveref}
\usepackage{subfig}
\usepackage[normalem]{ulem}

\begin{document}
\title{On Critical Temperature and Finite Size Scaling of Continuous Spin $2d$ Ising Model }

\author{Swapna Mahapatra}
\email{swapna.mahapatra@gmail.com}
\affiliation{\small  Department of Physics, Utkal University, Bhubaneswar-751004, India}
\author{Rudra Majhi}
\email{rudra.majhi95@gmail.com}
\affiliation{\small Department of Physics, Nabarangpur College, Nabarangpur-764059, India}
\author{Jahangir Mohammed}
\email{jahangirmd.physics@gmail.com}
\affiliation{\small Department of Physics, Nabarangpur College, Nabarangpur-764059, India}
\author{Subhashree Mohanty}
\email{subhaphy1@gmail.com}
\affiliation{\small Faculty of Science, Sri Sri University, Cuttack-754006, India}
\author{Priyanka Priyadarshini Pruseth}
\email{pkpruseth@gmail.com}
\affiliation{\small Department of Physics, Vesaj Patel College, Sundargarh-770076, India}
\author{Masoom Singh}
\email{masoom$_$singh@tamu.edu}
\affiliation{\small Department of Physics and Astronomy, Texas A$\&$M University, 578 University Dr, College Station, 77840, TX, USA}

\begin{abstract}
In this paper, we have studied a two-dimensional continuous-spin Ising model
in which the spin variables are bounded within the interval $s \in [-1,1]$.
We consider square lattice configuration with periodic boundary condition, 
no external magnetic field and no explicit on-site potential. Using Monte Carlo simulations, we show the existence of a second order phase transition with a critical temperature $T_c\approx$ $0.925$ J/$\text{k}_{B}$. We have also performed the Binder cumulant analysis
for estimating the critical temperature in our model and we find that the estimated value is consistent with the above value of $T_c$.
From the finite size scaling analysis, we obtain the critical
exponents associated with susceptibility, specific heat, magnetization, Binder cumulant and
we find that these values are in good agreement with the corresponding
values obtained for the two-dimensional Ising universality class. 
Finite-size scaling analysis is further supported by the scaling-collapse plots for susceptibility, magnetization and fourth-order Binder cumulant. 
The present formulation, based on the boundedness of the spin 
variables differs from the conventional soft-spin models where a local
potential is required to stabilize the system with unbounded spin amplitudes.  
Our findings show that boundedness alone is sufficient to ensure a stable Hamiltonian and exhibit a finite temperature phase transition with critical behaviour consistent with the $2d$ Ising Universality class.    
\end{abstract}
\maketitle



\section{Introduction}
The study of magnetism and its effect has been one of the fundamental
and exciting area of research for decades. It is a well known phenomenon
that in the absence of any
external magnetic field, when certain materials are cooled
down below a critical temperature called Curie temperature, spontaneous
magnetization appears giving rise to ferromagnetism.
Ising model is the simplest statistical model which has been
used to describe ferromagnetism and the model is analytically exactly
solvable in one and two dimensions. The two dimensional ($2d$)
square lattice Ising model consisting of discrete spin variables
$s_i = \pm 1$, exhibits a second order phase transition
at a critical temperature $T_c = 2.269$ J/$\text{k}_{B}$
\cite{Onsager:1944}.
In the standard Ising model, the Hamiltonian is given by,
\begin{equation}
H = -J\sum_{<ij>}s_i s_j - \mu \sum_j h_j s_j \;,
\end{equation}
where the first sum is taken over the nearest neighbours and
$J > 0$ (for ferromagnetic interaction) is the strength of the nearest
neighbour interaction, $s_i$ (which is a discrete
variable) represents the spin of the lattice site,
where each spin takes values $\pm 1$.
In the second term, $h_j$ is the external magnetic field at site $j$ and
$\mu$ is the magnetic moment.
Finding an exact analytic solution of the physical systems is very
difficult. The physics described by the standard Ising
model is quite nontrivial even though the Hamiltonian looks simple.
It exhibits the essential characteristics of
phase transition and critical phenomena, important features of many-body
problems such as long range correlation from the interaction of
nearest neighbours.
Numerical methods like Monte-Carlo simulations
are often used to study the critical behaviour of physical quantities, such as,
average energy $E$, magnetization $M$, susceptibility $\chi$ and
specific heat $C_v$. 

In the mean field approximation,
one replaces the nearest neighbour interactions by an average
field.
This means that the approximation is less affected by the dimensionality or
lattice configuration of the system.
In the $2d$ square lattice case, there are four nearest
neighbours and one can compute the critical temperature $T_c$ which
comes out to be
higher than the value obtained numerically
in Monte-Carlo simulations. This is because the mean-field approximation
ignores the fluctuations, which corresponds to deviations of the physical
quantities like energy, spin, magnetization from their average values due
to thermal motion. These fluctuations become large and long range
near the critical temperature $T_c$, which are not negligible in $2d$ models.
In the mean field approximation, every spin feels only the average
magnetization and the correlations between spins at different sites
are not taken into account.
However, in reality in $2d$ models, the effect
of correlated domain of spins due to fluctuations can not be ignored and
due to which the transition temperature obtained through Monte-Carlo
simulation turns out to be less compared to mean field approximation.

A natural generalization of the standard Ising model is to
consider the continuous spin Ising model where
spins take any value in a continuous interval.
In one of the early work by
Griffiths \cite{Griffiths:1969}, the standard Ising model with spin
$\frac{1}{2}\hbar $ has been extended to arbitrary spin
$\frac{p}{2}\hbar $, where an Ising particle of spin $\frac{p}{2}\hbar$ is
represented in terms of a cluster of $p$ spin $\frac{1}{2}\hbar$ particles
interacting with each other through ferromagnetic pair interactions.
Continuous spin Ising models have been studied extensively
in the context of soft-spin or $\phi^4 $
theories, where the spins take values in $\mathbb{R}$
and are stabilized by an on-site potential $V (s_i)$ which
basically controls the spin distribution.
The corresponding Hamiltonian is given by (with a vanishing external field),

\begin{equation}
H = -J\sum_{<ij>}s_i s_j + \sum_i V(s_i),
\end{equation}
where $J$ is the nearest neighbour interaction strength.
It is important to note that  continuous spin models
are known to exhibit critical behaviour, where the critical temperature
$T_c$ depends on the range of allowed spins 
and the form of the on-site potential $V(s_i)$. In these models, the on-site potential $V(s_i)$ plays an important role in
ensuring stability by controlling the amplitude fluctuations.
Continuous spin $2d$ Ising model has been studied
in various forms.
Analytic treatment with unbounded Gaussian spin model (where
spins can take any value in the interval $s_i \in [-\infty, +\infty]$)
and no on-site potential, has been explored to some extent predicting
$T_c = 0$. This implies the absence of a finite temperature phase transition,
which reflects the dominance of fluctuations. Some earlier discussions in the literature
have taken a nonzero potential $V(s_i)$ which
ensures $\mathbb{Z}_2 $ symmetry, bounded energy and staying within the Ising
universality class.
$V(s_i)$ is typically chosen as a harmonic potential or quartic
potential with $V(s_i) = a (s_i^2 -1)^2$, where the constant $a > 0$
controls the stiffness, meaning
for large $a$, the spins are tightly confined near $\pm 1$ and for
small $a$, spins are soft taking values away from $\pm 1$.
One may also consider the more familiar form of the quartic
potential corresponding
to $V(s_i) = a s_i^2 + b s_i^4$, where $a < 0$, $b >0$.
This double-well potential connects to $\phi^4$
field theory. It is known that the critical point of $\phi^4$ theory
is described by $2d$ Ising model \citep{Baker:1979}, \citep{Baker:1981}. See ref. \citep{Zinn-Justin:2007}, \citep{Parisi:1988} for field theoretical treatment of critical phenomena and Ising universality class.
Ref.\cite{Baker:1984} discusses limitations on universality in the continuous spin Ising model.
The effect of long-range interactions on the critical behaviour of
the continuous spin Ising model using Monte Carlo technique
has been discussed in \cite{Bayong:1999}, \citep{Bayongetal:1999}
where a phase transition has been observed
with nonstandard critical exponents. Percolation of clusters and
magnetization transition in classical continuous spin Ising model on
${{\mathbb{Z}}}_2$ have also been studied. Using Wolff cluster algorithm, Bialas
$\it {et~al.}$ \cite{Bialas:2000} have  obtained the critical point of the
percolation transition 
with $\beta_{crit} \sim 1.09312 $ (where, $\beta=J/k_B T$). Employing the method discussed in ref. \cite{Binder:1988}
they have shown that the percolation results for the critical
point matches with the thermal results and the values obtained for the
critical exponents are within the Ising universality class.
As an interesting application, the dynamics of
sea ice/water transition in Arctic region has been studied using bounded
continuous spin Ising model through numerical simulation \cite{Wang:2024}. 
We would like to draw the 
attention of the readers to the editorial article \cite{100-Glorious:2026}
``100 Glorious Years of the Ising Model'', which gives a very nice overview with  chronological milestones in the subject; focuses on the applications of Ising model as well as the latest research on Ising model and the associated results obtained by several researchers.

In this work, we consider a bounded continuous-spin Ising model in two
dimensional square lattice system, in which the
spin variables take continuous values in the interval $s_i \in [-1,1] $,
and no explicit on-site potential $V(s)$ is introduced.
This limit has not been systematically investigated in the literature in the context of critical behaviour. 
Here, we do not consider the external magnetic field.
In this formulation of bounded continuous spin system, we basically
want to ask the question that whether it is necessary to have an explicit
on-site potential for the emergence of a finite temperature phase transition with critical behaviour consistent with the 2d Ising universality class.
Using Monte Carlo simulations, we show
that this minimal model exhibits a well-defined second-order phase
transition at a critical temperature
$T_c \approx 0.925 \,$ (in the unit where $J=1=k_B$). 
We have also performed a Binder cumulant analysis of our model to estimate the critical temperature and we obtain $T_c \approx 0.925 \,$, which is consistent with the value obtained from the analysis of locations of peaks in the thermodynamic variables. From the finite-size 
scaling (FSS) analysis, we have obtained the associated critical
exponents and find that the values are consistent with the two-dimensional Ising universality class. 
We have also estimated the value of the correlation length critical exponent $\nu$ from the Binder cumulant data by considering a log-log plot of the derivative of the fourth-order Binder cumulant with respect to the temperature $T$ (at $T=T_c$) against the lattice size $L$.
We have investigated the finite size scaling behaviour of susceptibility, magnetization and fourth order Binder cumulant using the corresponding scaling-collapse plots as 
further evidence for the consistency of our results with the $2d$ Ising universality class.

Our study demonstrates that boundedness alone is sufficient to stabilize a finite temperature phase transition in a continuous-spin
system, even in the absence of an explicit on-site potential.
In other words, the bounded domain itself provides a natural constraint
on spin amplitudes and thereby removes the need for an additional
stabilizing potential.
This provides a minimal framework that interpolates between the Gaussian
model and the discrete Ising model. 
A vanishing on-site potential $V(s)$ corresponds to a model with uniform spin distribution and only nearest neighbour interactions. Our findings highlight the role of boundedness
in continuous-spin systems and we are able to demonstrate that amplitude
constraints alone can stabilize a finite-temperature phase transition
without the introduction of an on-site potential.
To our knowledge, this model for the $2d$ continuous spin Ising model has
not been explored before and our results are new. 

We would like to mention that the bounded continuous spin $2d$ Ising model ($s_i \in [-1,+1]$) can be viewed as statistically equivalent to the continuous, symmetric Heisenberg spin ($S=1$) model with the interaction restricted to the $z$ component only with Hamiltonian,
\begin{equation}
H=-J\sum_{\langle ij\rangle}S_i^zS_j^z .
\end{equation}
In terms of angular variables ($S_i^x = \sin\theta_i \cos\phi_i,
S_i^y=\sin\theta_i \sin\phi_i, S_i^z=\cos\theta_i$), the 
above Hamiltonian is given by,
\begin{equation}
H=-J\sum_{\langle ij\rangle}\cos\theta_i\cos\theta_j\,\,,
\end{equation}
which is independent of the azimuthal angles $\phi_i$. Identifying the spin variables in both the models, namely, $s_i=S_i^z=\cos\theta_i$, the resulting partition functions in both the models turn out to be the same except for a normalization factor of $(2\pi)^N$ (the factor of $2\pi$ arising from the integration over the azimuthal angles and $N$ is the 
number of lattice sites in the system), which is a constant. Both the models have discrete $Z_2$ symmetry and exhibit critical behaviour. It 
is worth mentioning that the standard classical, isotropic Heisenberg model with unit vector spins with the Hamiltonian
$H_H= -J\sum_{\langle ij\rangle}(S_i^x S_j^x + S_i^y S_j^y + S_i^zS_j^z)$
is invariant under continuous $O(3)$ rotational symmetry 
and the model in $2d$ with short range interactions does not exhibit a conventional finite temperature order-disorder phase transition \cite{Mermin-Wagner:1966}. The interval of the bounded, continuous spin $s_i$ $\in [-1, +1]$ in our model can be understood from the following consideration: writing $S^z$
(in the unit vector spin Heisenberg model with interaction restricted to $S^z$ component only) in terms of angular variable, namely, $S_i^z = \cos\theta_i$ gives the range $-1 \le \cos\theta_i \le +1$ because of the geometry of the unit sphere. Therefore, the range of the continuous spin $s_i \in [-1, +1]$ considered in our model, arises naturally when $s_i$ is identified with the $z$ component of the restricted classical, 
continuous unit vector spin Heisenberg model with the Hamiltonian involving only $S_i^z$. 

It is interesting to note that the bounded continuous spin $2d$ Ising model considered here, is similar to kinetic exchange model of opinion formation in a society proposed by Biswas-Chatterjee-Sen (BChS model) \cite{BChS:2012}, \citep{ BChSMC:2023}. In the above model, the nature of the disorder induced phase transition, critical behaviour have been investigated and it has been shown that the critical exponents for the models with both continuous as well as discrete opinions belong to the same Ising universality class.

This paper is organised as follows: in section 1, we have given a brief introduction to the subject of $2d$ Ising model with emphasis on
continuous spin Ising model along with a discussion on on-site potential. We also explain our approach which is different from the conventional
set-up, highlighting the important findings in the context of finite temperature phase transition and critical behaviour in the present formulation. In section 2, we describe the details of the numerical simulation which we have used to study the critical behaviour of the bounded continuous spin $2d$ Ising model. In particular, we have studied the behaviour of Susceptibility ($\chi$), Specific heat ($C_v$) and Magnetization ($M$) for obtaining the critical temperature $T_c$ as well as the critical exponents. Section 3
describes our results and analysis, where subsection 3.1 deals with the
critical temperature obtained for our continuous bounded spin $2d$ Ising model 
from the analysis of the locations of the peaks in the above thermodynamic variables; we present the Binder Cumulant analysis for our model in subsection 3.2 which gives an independent estimation of the critical temperature $T_c$; 
subsection 3.3 describes the role of amplitude fluctuations; subsection 3.4 deals with the finite-size scaling and
critical exponents in our model, which we compare with the values for $2d$ Ising
universality class. We also study the finite-size scaling behavior of susceptibility, magnetization and Binder cumulant through the corresponding scaling-collapse plots.
We present the summary and perspective of our work in section 4.
\section{Numerical Simulation}
We investigate the thermodynamic properties of the $2d$ ferromagnetic Ising model on square lattices using Monte Carlo simulations based on the Metropolis algorithm~\cite{Metropolis:1953}. The system is described by the Hamiltonian \begin{equation}
\label{eq:eqn3}
 H(s) = - \frac{J}{2}\sum\limits_{i,j=1}^{L} s_{i,j} (s_{i,j-1}+s_{i,j+1}+s_{i-1,j}+s_{i+1,j})
\end{equation}
where $ J > 0$ denotes the ferromagnetic exchange coupling constant (set to unity throughout) (the factor of $\frac{1}{2}$ is introduced to prevent double counting of the interaction energy between pairs of neighbouring spins) and $s_{i,j}$ is the spin variable at lattice site $i,j$. In our implementation, spins are continuous variables constrained to the interval $s_{i,j} \in [-1, +1]$, representing a continuous-spin generalization that bridges the classical discrete Ising model and continuous vector spin models. This formulation enables exploration of spin magnitude fluctuations and their role in magnetic ordering phenomena. 
We employ periodic boundary condition wherein spins at position $(i, j)$ interact with neighbours at positions $(i \pm 1 \mod L, j)$ and $(i, j \pm 1 \mod L)$ for a lattice of linear dimension $L$.

The simulation procedure employs the standard Metropolis algorithm to sample configurations from the canonical ensemble at temperature $T$. Each Monte Carlo sweep consists of $L^2$ attempted single-spin updates, ensuring that on average every lattice site is visited once per sweep. For each update attempt, a lattice site $(i,j)$ is selected uniformly at random, and a new spin value $s_{\text{new}}$ is proposed by drawing from a uniform distribution over the allowed spin range: $s_{\text{new}} = 2r - 1$ where $r \sim (0,1)$. In this approach, each proposed spin is completely independent of the current spin value, in contrast to local perturbative updates where new spins are generated as $s_{\text{new}} = s_{\text{old}} + \delta s$ with small random perturbations $\delta s$.  The energy change associated with the proposed spin flip is computed as $\Delta E = -J(s_{\text{new}} - s_{\text{old}}) \sum_{\text{neighbours}} s_{\text{neighbours}}$ using the current spin configuration. The proposed move is accepted with probability $\min(1, \exp(-\beta \Delta E))$ according to the Metropolis criterion, which satisfies detailed balance~\cite{Landau:2014} and asymptotically samples the Boltzmann distribution. To maintain numerical efficiency, we employ incremental energy updates wherein the total system energy is updated by adding $\Delta E$ for accepted moves rather than recalculating the entire lattice energy at each step.

The number of Monte Carlo sweeps is scaled with the lattice size to account for the
longer relaxation times of larger systems. One sweep consists of $L^2$ attempted
single-spin updates. For a lattice of linear size $L$, each run consists of
$N_{\rm sweeps}=50{,}000\times\max(1,L/8)$ sweeps, i.e.\ from $5\times10^{4}$ sweeps for
$L=8$ to $8\times10^{5}$ sweeps for $L=128$. 
The first 90\% of each
run is discarded as equilibration, and observables are measured after every sweep
during the remaining 10\%. The same schedule is used at all temperatures.

Simulations cover the range $T\in[0.1,\,2.3]$ (in units where $J=1=k_B$). Away from the
transition we use a coarse grid, $\Delta T=0.2$ for $T>1.3$, while the region around
the transition, $0.1\le T\le1.3$, is sampled with $\Delta T=0.025$. For the Binder
cumulant analysis,\,(Sec.~\ref{sec:BC}) we consider a finer grid of $\Delta T=0.0125$ in $0.85\le T\le1.00$. The critical window was chosen from a preliminary coarse scan, in which
the peaks of $C_v$ and $\chi$ appear near $T\approx0.925$. This concentrates computational
resources on the critical region while still covering the full phase diagram.

The primary thermodynamic observables computed from our simulations include the total magnetization per site, defined as $M = L^{-2} \sum_{i,j} s_{i,j}$, which serves as the order parameter for ferromagnetic ordering. The energy per site is calculated as $E = H/L^2$ where the total energy is obtained by summing nearest-neighbour interactions with care taken to avoid double-counting. Specific heat and magnetic susceptibility are derived from fluctuations via the fluctuation-dissipation theorem. The specific heat per site is given by 

\begin{equation}
  C_{v} = \beta^2 L^{-2} [ \langle E^2 \rangle - \langle E \rangle^2 ]  
\end{equation}
where $\beta = (k_B T)^{-1}$ is the inverse temperature with Boltzmann constant $k_B$ set to unity, and $E$ denotes the total system energy. The magnetic susceptibility is computed as 
\begin{equation}
\chi = \beta L^2 [ \langle M^2 \rangle - \langle M \rangle^2 ],    
\end{equation}
 with angular brackets denoting thermal averages over Monte Carlo samples, which exhibits pronounced peaks at the critical temperature.

The Curie temperature $T_c$ characterizing the ferromagnetic phase transition is determined through multiple complementary criteria. We identify $T_c$ from the temperature at which the specific heat attains its maximum value, from the peak in magnetic susceptibility, and from the maximum slope of the magnetization curve. The reported critical temperature represents the average of these estimates, providing a robust determination that accounts for finite-size effects and statistical uncertainties. Finite-size scaling analysis is performed by simulating square lattices with lattice sizes $L = 8, 16, 32, 64$, $128$ and corresponding critical exponents have been extracted.

Data analysis procedures include careful treatment of autocorrelation effects to ensure statistically independent samples. While we do not explicitly compute autocorrelation times, the long equilibration periods and substantial number of post-equilibration sweeps ensure that measurements are effectively decorrelated. For each temperature and run, we accumulate measurements of energy and magnetization during the production phase at each Monte Carlo sweep, from which we compute temporal averages and second moments needed for fluctuation-based quantities. Specifically, we store $E$ and $E^2$ for each measurement, then calculate $\langle E \rangle = n_{\text{meas}}^{-1} \sum_{T} E(T)$ and $\langle E^2 \rangle = n_{\text{meas}}^{-1} \sum_{T} E^2(T)$ where $n_{\text{meas}}$ is the number of measurements after equilibration. These run-level statistics are then combined across independent runs to yield final ensemble averages and uncertainties. 

The computational implementation achieves efficiency through several algorithmic optimizations. Energy updates are performed incrementally by calculating the change $\Delta E$ associated with each proposed spin flip and updating the total energy accordingly, avoiding the computational expense of recalculating the full lattice energy after every move. Random number generation employs efficient algorithms to minimize overhead. For large lattices approaching the thermodynamic limit, these optimizations enable practical simulation times while maintaining numerical accuracy and physical fidelity.

The simulation which we have followed here, differs from several established methods in the literature in significant ways. While cluster algorithms such as Swendsen-Wang~\cite{Swendsen:1987} and Wolff~\cite{Wolff:1989} offer superior performance near criticality by dramatically reducing autocorrelation times through collective spin flips, these methods are primarily designed for discrete spin systems and their extension to continuous-spin models presents considerable algorithmic complexity (see ref.\cite{Herrmann:2021} for a modern perspective on computational statistical physics). The conventional approach for continuous-spin systems typically employs local perturbative updates where proposed spins are generated as $s_{\text{new}} = s_{\text{old}} + \delta s$ with small random perturbations, following the methodology commonly used in classical $XY$ and Heisenberg models. In contrast, our global spin proposal scheme generates completely independent spin values uniformly distributed over $[-1, +1]$, which ensures ergodicity and prevents the system from becoming trapped in metastable configurations that can plague the local update schemes. Heat bath algorithms provide an alternative to Metropolis acceptance by directly sampling from the conditional distribution, but for continuous spins with nearest-neighbour interactions, the conditional distribution lacks a simple closed form, necessitating rejection sampling that offers little advantage over our approach. Advanced techniques such as parallel tempering or population annealing~\cite{Machta:2010} could potentially improve the efficiency of our simulations, but these methods are significantly more difficult to implement. These techniques are primarily designed to help systems escape from incorrect configurations where the system appears stable but has not reached the true equilibrium state. Such incorrect stable states are common in complex magnetic systems like spin glasses, where magnetic interactions compete with each other and create many nearly equivalent configurations. However, in the ferromagnetic Ising model studied here, spins prefer to align with their neighbours, leading to a single clear ground state, so these advanced sampling techniques offer little practical advantage for our purposes. Here, adaptive two-phase temperature sampling strategy saves computational time by focusing
measurements near the critical region. We first perform a coarse temperature scan to locate the
approximate critical temperature, then perform fine sampling in that region. This approach is more efficient than sampling all temperatures uniformly with fine resolution.This methodology provides a comprehensive framework for investigating ferromagnetic phase transitions in two-dimensional Ising model. The combination of continuous spin variables, adaptive sampling strategies, careful equilibration protocols, and systematic finite-size analysis enables approximate determination of critical temperatures and characterization of thermodynamic properties across the full phase diagram from the ordered ferromagnetic state at low temperatures through the critical region to the paramagnetic phase at high temperatures.

\section{Results and Discussion}
\subsection{Critical Temperature} 
In this section, we discuss the results of Monte-Carlo simulations. To study the $T_c$ of continuous bounded spin Ising model, we have taken different square lattice sizes upto $128\times 128$ and temperature ranging from $0.1$ to $2.3$ with periodic boundary condition. 

Fig.~\ref{fig:Tvschi_m_e_cv} is the simulation for $64\times64$ lattice size and it shows the variation of the average magnetization per site $M$, 
specific heat per site $C_v$, magnetic susceptibility per site $\chi$, and energy per site as functions of temperature. The colored circles represent the computed values of the corresponding observables at different temperatures. From Fig.~\ref{fig:Tvschi_m_e_cv}, one finds that the energy gradually increases to zero and magnetization gradually decreases to zero. Susceptibility and specific heat also change, initially they increase upto $T_{c}$ and then start decreasing as shown in Fig.~\ref{fig:Tvschi_m_e_cv}. From the $\chi$ vs $T$ and $C_v$ vs $T$ graphs, one can observe a peak giving a clear signature of the critical temperature $T_c \sim 0.925$ (in the units where $J=1=k_B$). This indicates that the system undergoes a second order phase transition at $T_c$ around  $ 0.925$. 

In a finite system like our bounded continuous spin model with no on-site
potential, we find that magnetization $M$ decreases smoothly from a
nonzero value as the temperature $T$ increases. The temperature where $M$ seems to vanish, depends
on the lattice size $L$ and it shows rounding near the transition without a
sharp discontinuity due to finite size effects. Susceptibility
$\chi$ exhibits a prominent peak at a temperature close to the critical
temperature $T_c$, which is found to sharpen and shifts slightly when we
increase the lattice size $L$. We also observe that the graph for specific heat $C_v$ has a dominant peak for each $L$ indicating the
critical behaviour.
\begin{figure}[htbp!]
    \centering
    \includegraphics[width=1\linewidth]{plot/ising_64x64_observables_no_interpolation.png}
\caption{Temperature evolution of  magnetization $M$,  specific heat $C_v$, susceptibility $\chi$, and energy per site $E$ for a $64 \times 64$ lattice, showing the characteristic signatures of the ferromagnetic-paramagnetic phase transition.}
    \label{fig:Tvschi_m_e_cv}
\end{figure}

\subsection{Binder Cumulant Analysis}
\label{sec:BC}
\begin{figure}
    \centering
    \includegraphics[width=0.8\linewidth]{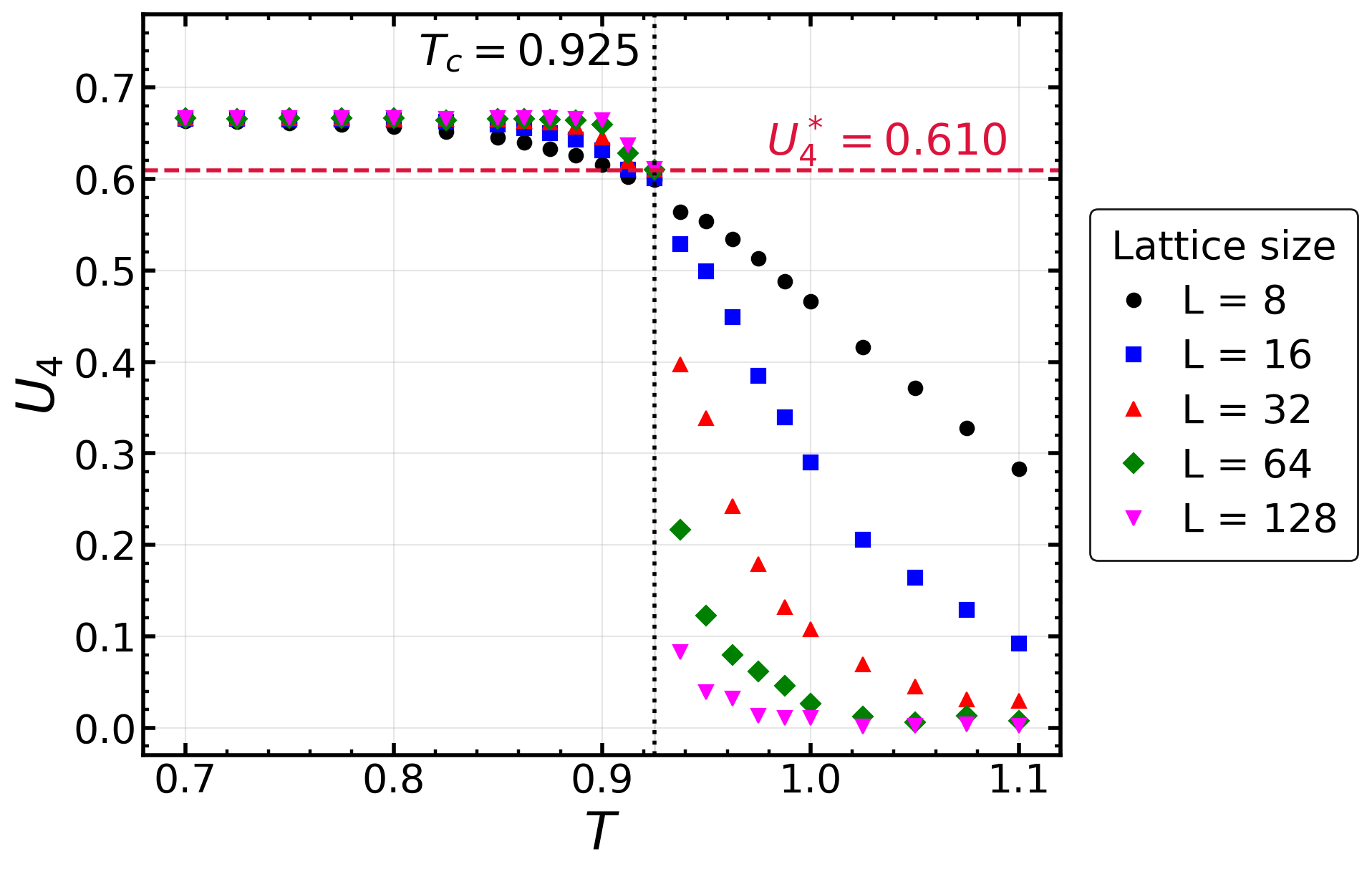}    
    \caption{Fourth-order Binder cumulant vs. temperature.}
    \label{fig:bindercumulant}
\end{figure}

Binder cumulant analysis offers an independent method for estimating the critical temperature $T_c$ for a finite size system
and helps in understanding how the system behaves as it goes through the
phase transition.
In this section, we study the fourth order Binder cumulant for different lattice sizes, 
where the crossing of the Binder cumulant curves for different lattice
sizes provides an estimate of the critical temperature $T_c$
\cite{Binder:1981}.
The fourth-order Binder cumulant is defined as,
\begin{equation}
U_4 = 1 - \frac{\langle M^4 \rangle}{3\langle M^2 \rangle^2},
\label{eq:binder_cumulant}
\end{equation}
where $M$ is the magnetization per spin and $\langle M^i \rangle$, with $i = 2$ and $4$, denotes the thermal average of $M^i$. Fig. \ref{fig:bindercumulant} shows $U_4$ as a function of temperature for different lattice sizes. We find that the curves for different lattice sizes cross approximately at $T = 0.92503 \pm 0.00003$ (jackknife) which we take as the critical temperature $T_c = 0.925$. 
The value of the critical temperature as obtained from the crossing of the Binder cumulant curves is consistent with the
value of $T_c$ which we have obtained from the analysis of the location of
the peaks in the thermodynamic variables through Monte Carlo simulations. 
At $T=T_c$, the value of the critical Binder cumulant $U_4^{*}$ (from Fig. \ref{fig:bindercumulant}) is found to be $0.6095\pm 0.0010$ (jackknife) which we take as $U_4^* = 0.610$. This value of $U_4^*$ is independent of the system size $L$ and is consistent with the established result.

\begin{figure}
    \centering
    \includegraphics[width=0.6\linewidth]{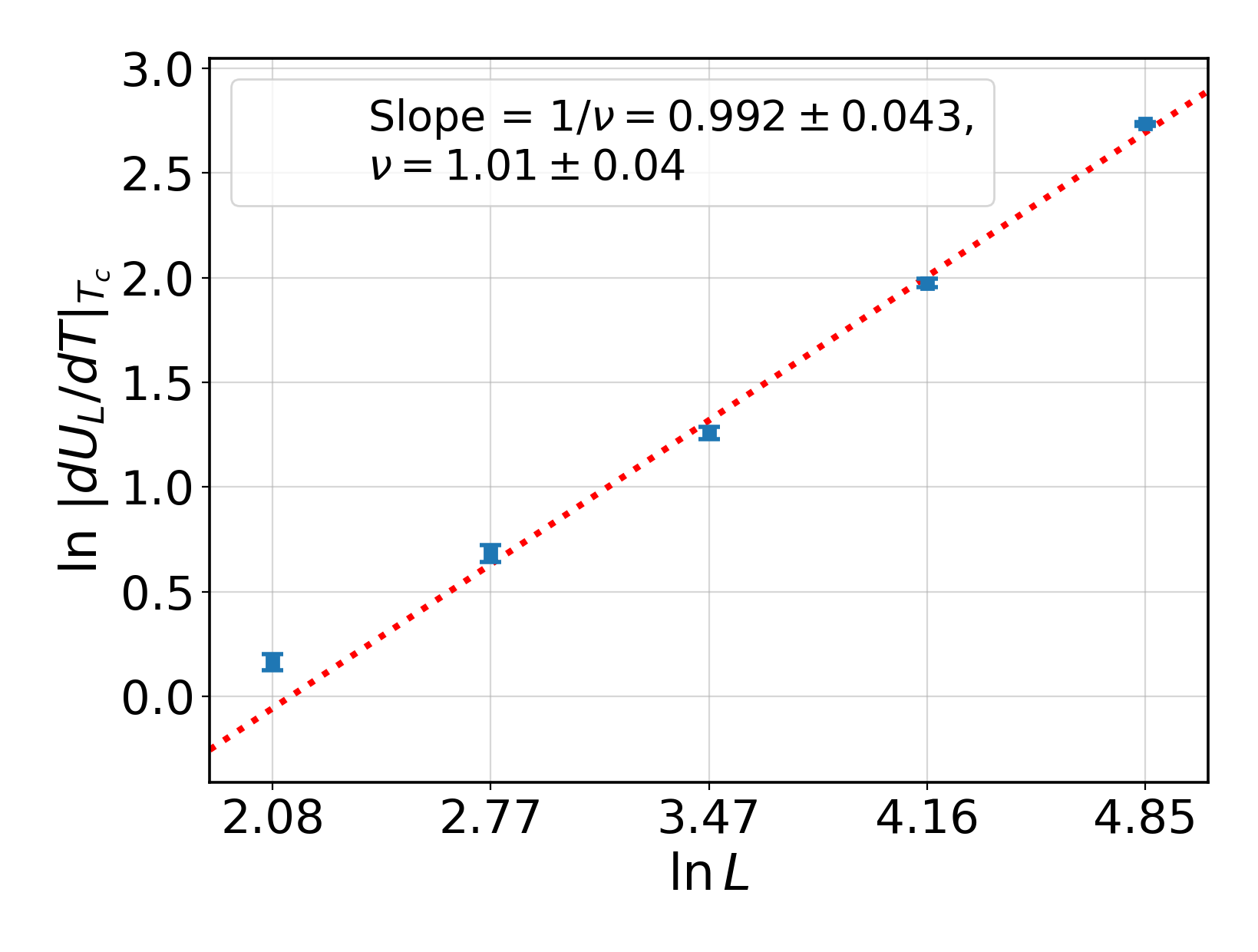}
    \caption{Finite-size scaling of the Binder cumulant slope with $T_c =0.925$, which gives $\nu = 1.01\pm 0.04$.}
    \label{fig:u4-derivative}
\end{figure}

Having obtained the critical temperature $T_c$, we then estimate the correlation length critical exponent $\nu$ which characterizes the divergence of the correlation length 
$\xi$ as the temperature approaches the critical temperature $T_c$. The 
relationship is given by, 
\begin{equation}
\xi \sim {\vert T - T_c \vert}^{-\nu}
\end{equation}
We extract the value of $\nu$ from the Binder cumulant data using the relation 
\begin{equation}
{\vert \frac{d U_4}{d T} \vert}_{T=T_c} \propto 
L^{1/{\nu}}
\end{equation}
We consider a log-log plot of the derivative of the Binder cumulant $U_4$ with respect to the temperature $T$ (at $T=T_c$) as a function of $L$, where $T_c$ is determined as above. According to finite-size scaling theory, the slope of this plot is given by $1/{\nu}$, from which we extract the value of $\nu$.
The plot is shown in Fig. \ref{fig:u4-derivative}, from which we obtain $\nu = 1.01 \pm 0.04$, which is in good agreement with the exact value ($\nu=1$) for the two-dimensional Ising universality class.
 \vspace{-1cm}
\subsection{ Amplitude fluctuation}
\begin{figure}
    \centering
    \includegraphics[width=0.6\linewidth]{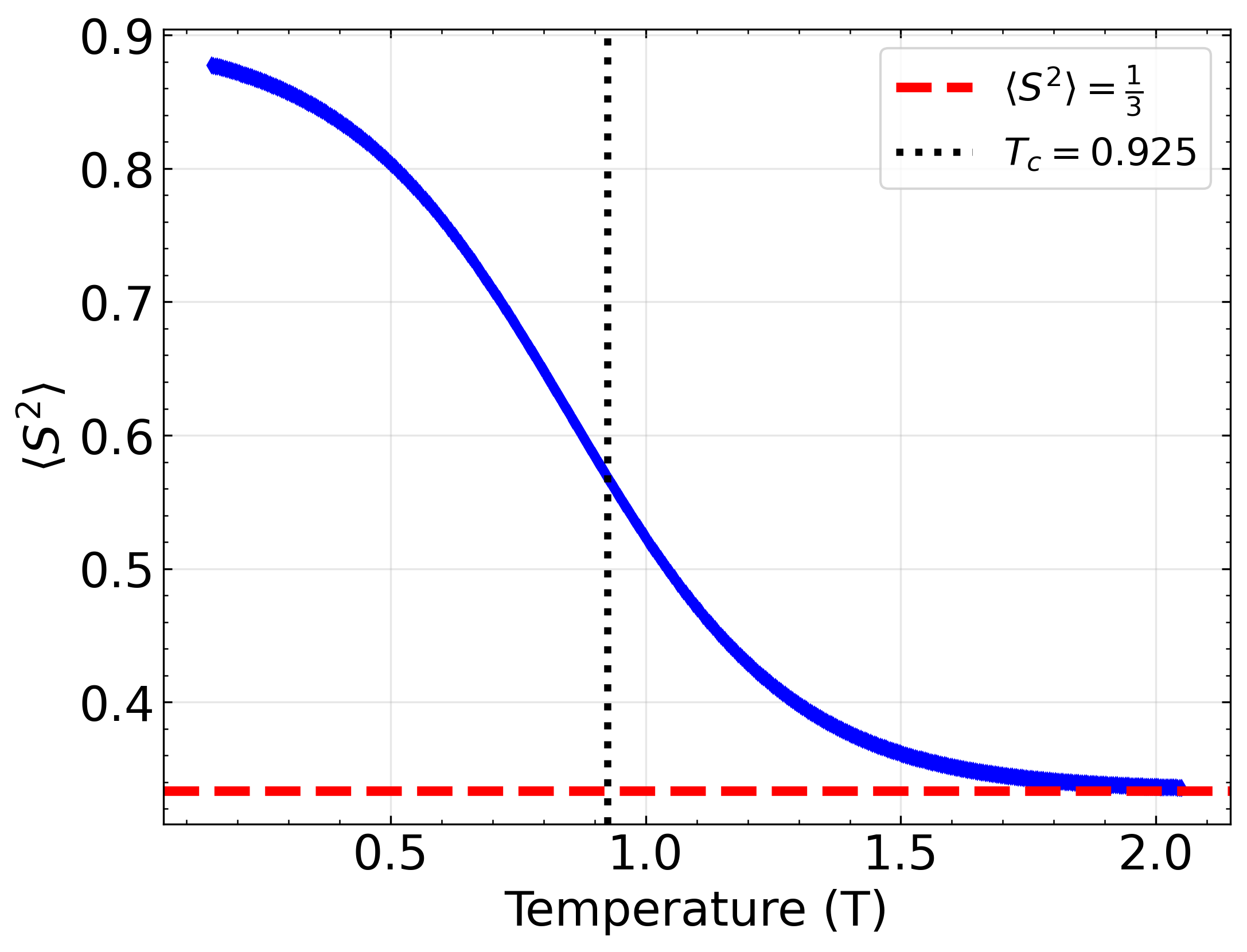}
    \caption{Mean square spin $\langle s^2 \rangle$ as a function of Temperature ($T$) for lattice size $L = 32$.}
    \label{fig:amplitude_flctuation}
\end{figure}

An important aspect of the present formulation is the role of amplitude
fluctuations of the spin variables. In the discrete Ising model
($s_i = \pm 1$), there is no amplitude fluctuation since the magnitude
of the spin is fixed. However, in the continuous spin model
considered here, the magnitude of the spins varies continuously
in the bounded interval $[-1,1]$, which leads to amplitude fluctuation. 
This is important as the interaction term in the Hamiltonian depends on the
product $s_i s_j$, which can be written as the product of the spin signs
multiplied by their magnitudes. Since the ordering depends on the signs
and amplitudes, the interaction strength or the effective coupling
becomes weaker for small fluctuations in the amplitude in the system
even if the spins are aligned.
One can qualitatively understand this from the effective interaction
picture, where $T_c \sim J \langle s^2 \rangle$ ($\langle s^2 \rangle$ is the mean-square spin and measures the typical spin amplitude).
Amplitude fluctuations are large at high temperature and are small at low temperature. 
At high temperature, thermal 
fluctuations dominate over the  
interaction energy and 
all spin configurations in the bounded domain $[-1,1]$ become almost equally probable. So for such uniform distribution in the interval $[-1,1]$, the mean-square spin yields 
$\langle s^2 \rangle \approx \frac{1}{3}$. As the temperature decreases, thermal fluctuations are suppressed and interaction energy becomes more important. This scenario favours spin configurations with minimum energy which pushes the spins towards the boundary $\pm 1$ and this corresponds to the mean-square spin value 
$\langle s^2 \rangle \approx 1$. At low temperature, all 
spin configurations have nearly the same amplitude as they 
cluster around $\pm 1$. Hence the amplitude fluctuations 
become much smaller. This implies that amplitude fluctuations are suppressed as the temperature decreases and the mean 
square spin $\langle s^2 \rangle $ increases from  $\frac{1}{3}$ at high temperature to $1$ at very low temperature.  
This qualitative behaviour of $\langle s^2 \rangle$ with respect to change in temperature is plotted 
in Fig. \ref{fig:amplitude_flctuation}. To analyze amplitude fluctuations, we simulated a lattice size of \(L = 32\) over a temperature range of T = 0.15 to 2.5 using periodic boundary condition. As shown in Fig. \ref{fig:amplitude_flctuation}, the curve reveals three distinct regions: the low temperature region ($0.15 < T < 0.6$) where fluctuations are highly restricted, showing a very narrow variance; transition region ($0.6 < T < 1.1$) where the curve drops sharply, signaling massive, scale-invariant fluctuations and a high temperature region ($T > 1.1$) where the curve flattens out perfectly along the horizontal dashed line (red) at $\langle s^2 \rangle \sim 0.3333$. This asymptotic behaviour proves that collective interactions break down at high temperature, which implies that the spins behave as uniformly distributed random numbers bounded between -1 and 1.
 The amplitude fluctuation in the bounded domain of spin
as considered here, plays a key role in reducing large fluctuations in the spin amplitude leading to a reduction of the effective interaction strength and a reduced critical temperature as compared to the discrete Ising model case. 
Our observed value of $T_c \sim 0.925 $ basically
reflects the finite but reduced amplitude fluctuations in the
continuous-spin system, where, the bounded domain provides a constraint
on amplitude fluctuations. This picture is very different from the discrete Ising case.
 \vspace{-0.5cm}

 \vspace{-0.1cm}

\subsection{Finite-size scaling analysis}
In order to study the phase transition and critical temperature
in our continuous bounded spin model further, we have considered finite size
scaling (FSS) and studied the
critical exponents. FSS framework
basically helps to understand the behaviour of the physical observables near
the critical point for finite system size.
In a finite system, the divergences in magnetization,
susceptibility, specific heat etc are smoothed.
Near the critical point, thermodynamic observables exhibit characteristic
power-law dependence on lattice size $L$ given by ~\cite{Fisher:1972}, \citep{Cardy:1996}, \citep{Stanley:1971},
\begin{equation}
        \chi_{\max}(L) \sim L^{\gamma/\nu}, \qquad
M(T_c, L) \sim L^{-\beta/\nu}, \qquad
C_v^{\max}(L) \sim C_0 \log L ,
\label{critical_exponenents}
\end{equation}
where $\gamma$, $\beta$, $C_0$ and 
$\nu$ are the critical exponents and $\nu =1$ for $2d$ Ising model. 

\begin{figure}[htbp!]
    \includegraphics[width=1\linewidth]{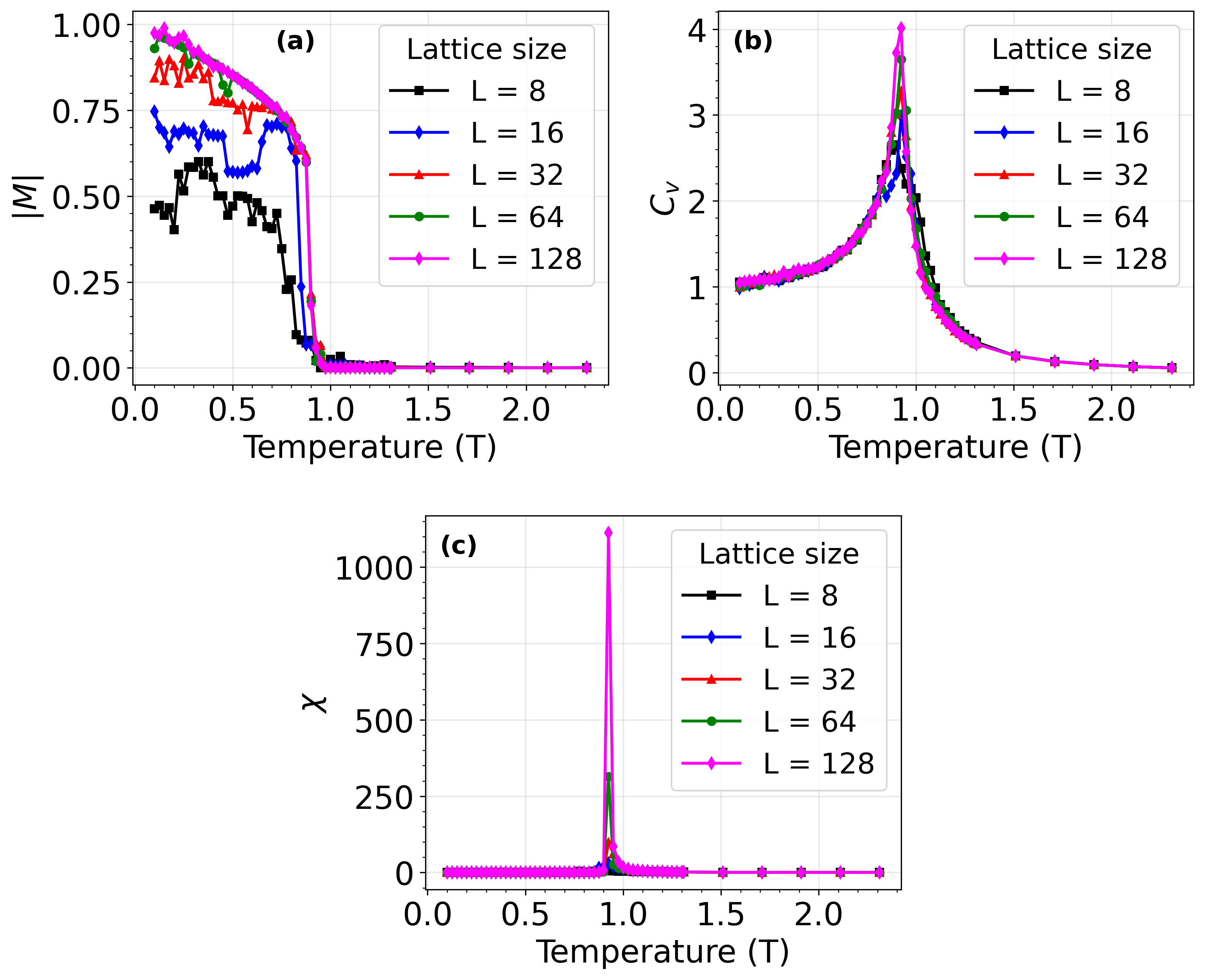}
\caption{(a) $|M|$ vs. $T$, (b) $C_v$ vs. $T$, and (c) $\chi$ vs. $T$ for lattice sizes $L=\mbox{8, 16, 32, 64, 128}$.}
    \label{fig:periodic}
\end{figure}

\begin{figure}[htbp!]
    \centering
    \includegraphics[width=1\linewidth]{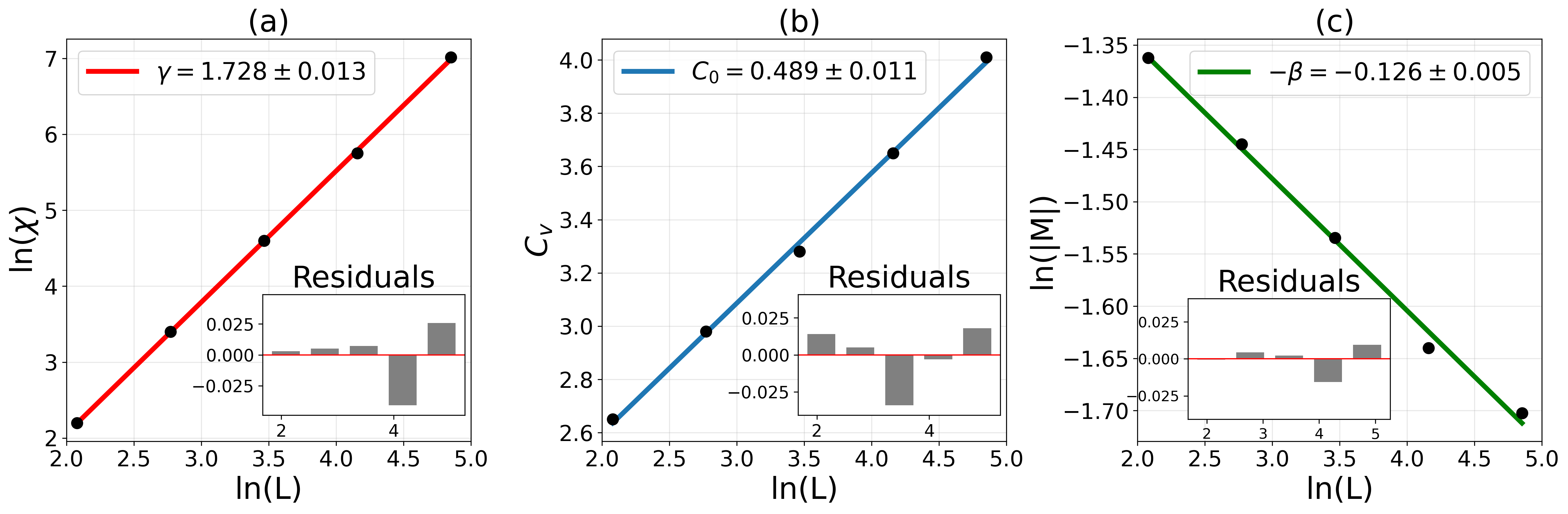}   
    \caption{Critical exponents characterizing the ferromagnetic phase transition under periodic boundary condition. Values of $\gamma$ (susceptibility exponent), $C_0$ (specific heat exponent) and $\beta$ (order parameter exponent) are obtained from finite-size scaling analysis.}
    \label{fig:Critical_exponent}
\end{figure}

By analyzing the size dependence of peak positions and magnitudes
in susceptibility and specific heat, we can extract the
critical exponents and verify consistency with the expected universality
class. We can now use the above equations~\ref{critical_exponenents} and determine their appropriate exponents. This is done by taking the peak values for the
collected data of the observable and plotting a $ ln-ln $ graph which
yields a straight line with gradient being equal to the respective critical exponent.
We have considered square lattice sizes 8, 16, 32, 64 and 128 to determine the critical exponents. We have estimated the peak values
of $|M|$, $C_v$ and $\chi$ by plotting $|M|$ vs. $T$, $C_v$ vs. $T$
and $\chi$ vs. $T$ in Fig.~\ref{fig:periodic}. The result for lattice size \(L=8\) in Fig.~\ref{fig:periodic} is poor due to the relatively small spin configuration. These estimations of the observables are used
to find the critical exponents.  The calculation for the critical exponents are shown in
Fig. \ref{fig:Critical_exponent} and the estimated values and theoretical
values of the critical exponents are listed in Table \ref{tab:critical_expo}.

Comparing with the values obtained
for standard $2d$ Ising model,
we find that our results for the critical exponents for the continuous bounded spin $2d$ Ising model are in good agreement with those
in the $2d$ Ising universality class.

\begin{figure}
    \centering
    \includegraphics[width=0.49\linewidth]{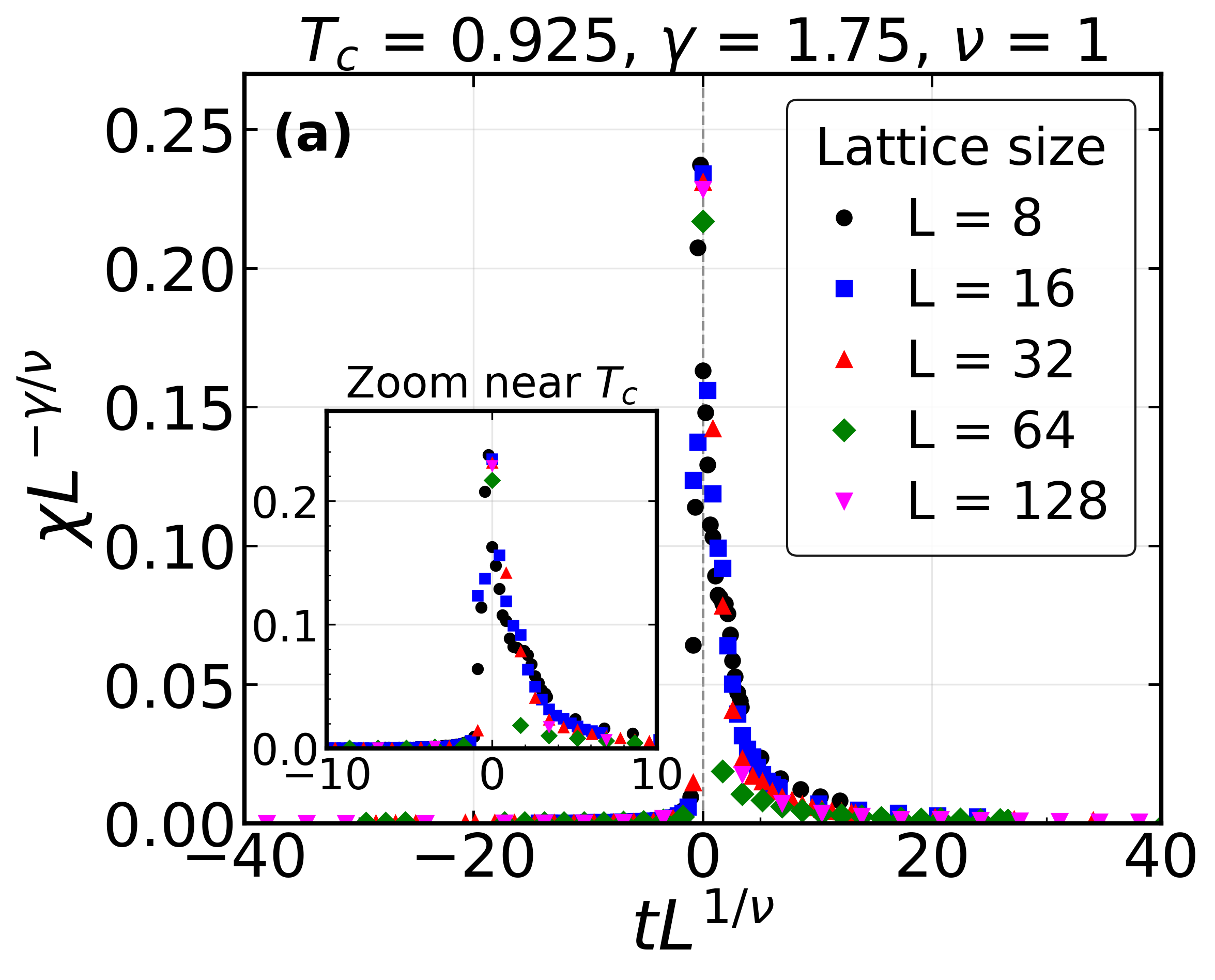}
    \includegraphics[width=0.49\linewidth]{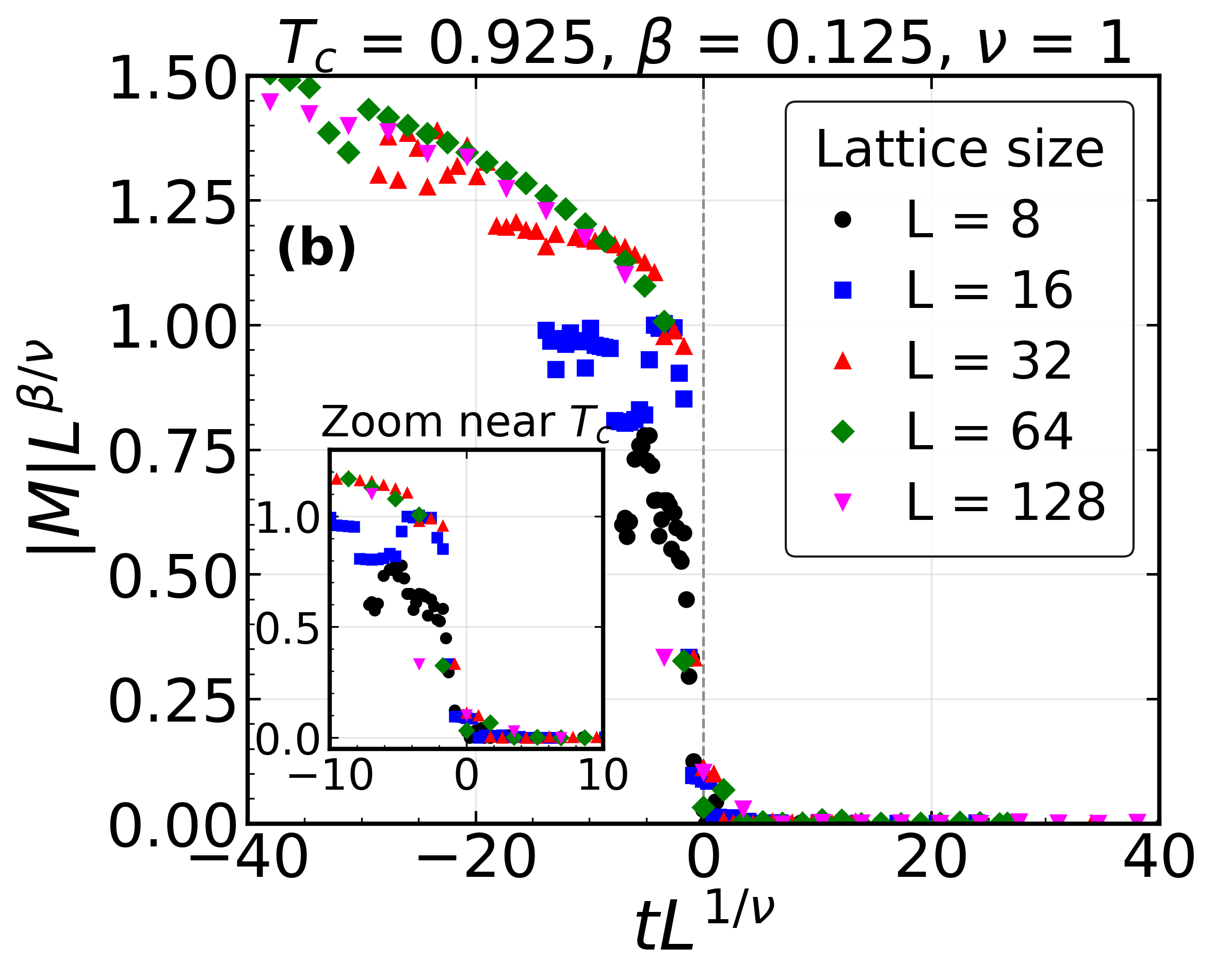}
    \includegraphics[width=0.49\linewidth]{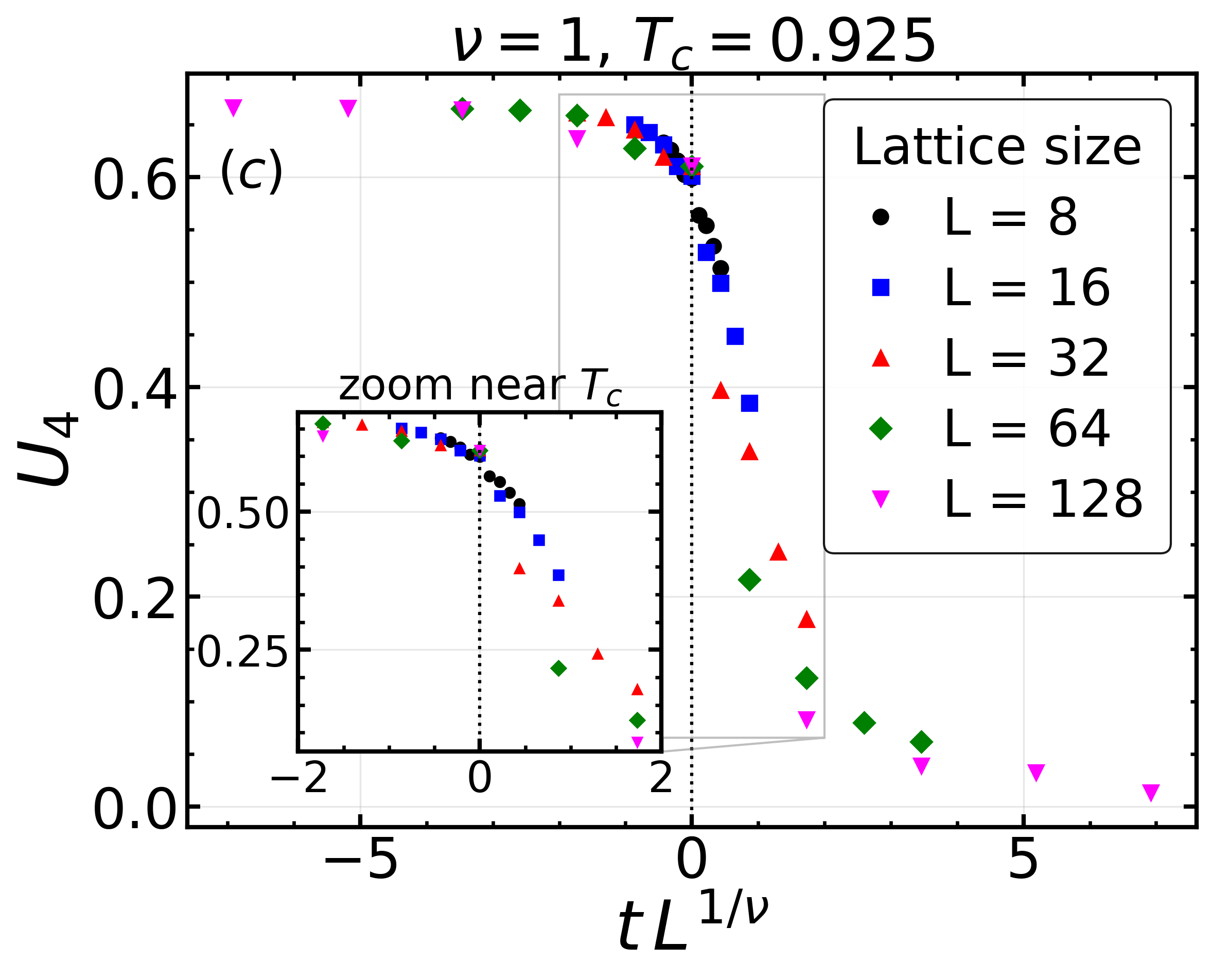}
    \caption{Scaling-collapse plots for $(a)$ Susceptibility, $(b)$ Magnetization and $(c)$ fourth-order Binder cumulant with $T_c=0.925$; $\nu=1$, $\gamma = 7/4$, and $\beta = 1/8$.}
    \label{fig:scaling-collapse}
\end{figure}

\begin{table}[ht]
\caption{Critical exponents for the two-dimensional continuous spin Ising model.\label{tab:critical_expo}}
{
\begin{tabular}{lccc}
\toprule
\textbf{Physical Observable} &
\textbf{Critical Exponent} &
\textbf{Estimated Value} &
\textbf{Theoretical Value} \\
\colrule
$\chi$  & $\gamma$ & $1.728 \pm 0.013$ & $1.75$ \\
$C_v$   & $C_0$    & $0.489 \pm 0.011$ & $0.50$ \\
$M$      & $\beta$  & $0.126 \pm 0.005$ & $0.125$ \\
\botrule
\end{tabular}
}
\end{table}
Next, we investigate the finite-size scaling behaviour of the susceptibility 
$\chi $, magnetization $M$ and fourth-order Binder cumulant $U_4$ by constructing the corresponding scaling-collapse plots. Near the critical temperature, finite-size scaling  predicts the following relations for $\chi$, $M$ and $U_4$, namely, 
\begin{eqnarray}
\chi(T,L)\,L^{-\gamma/\nu} =\mathcal{X}\left(tL^{1/\nu}\right) \\
\nonumber
\vert M(T,L) \vert \, L^{\beta/\nu} = \mathcal{M}\left(tL^{1/\nu}\right) \\
\nonumber
U_4(T,L)=\mathcal{U}_4\left(tL^{1/\nu}\right)
\end{eqnarray}
\noindent 
where, $\mathcal{X},\, \mathcal{M}$ and $\mathcal{U}_4$ are 
universal scaling functions in the above expressions and $t= \frac{T-T_c}{T_c}$. 
We study the finite size scaling behaviour for $\chi$, $M$ and $U_4$ by plotting \, 
$y = \chi {L^{\gamma/\nu}}$; $y=
\vert M \vert L^{\beta/\nu}$ and $y=U_4(L,T)$ against $x=t L^{1/\nu}$
respectively (where we have taken the exact value of $\nu=1$, $\beta = 1/8$, $\gamma = 7/4 $) as shown in Fig. \ref{fig:scaling-collapse}. 
We find that the curves for the rescaled variables for different lattice sizes $L$, collapse approximately onto a single curve for each of the above thermodynamic variables and Binder cumulant. 

\begin{figure}[htb!]
    \centering
    \includegraphics[width=0.49\linewidth]{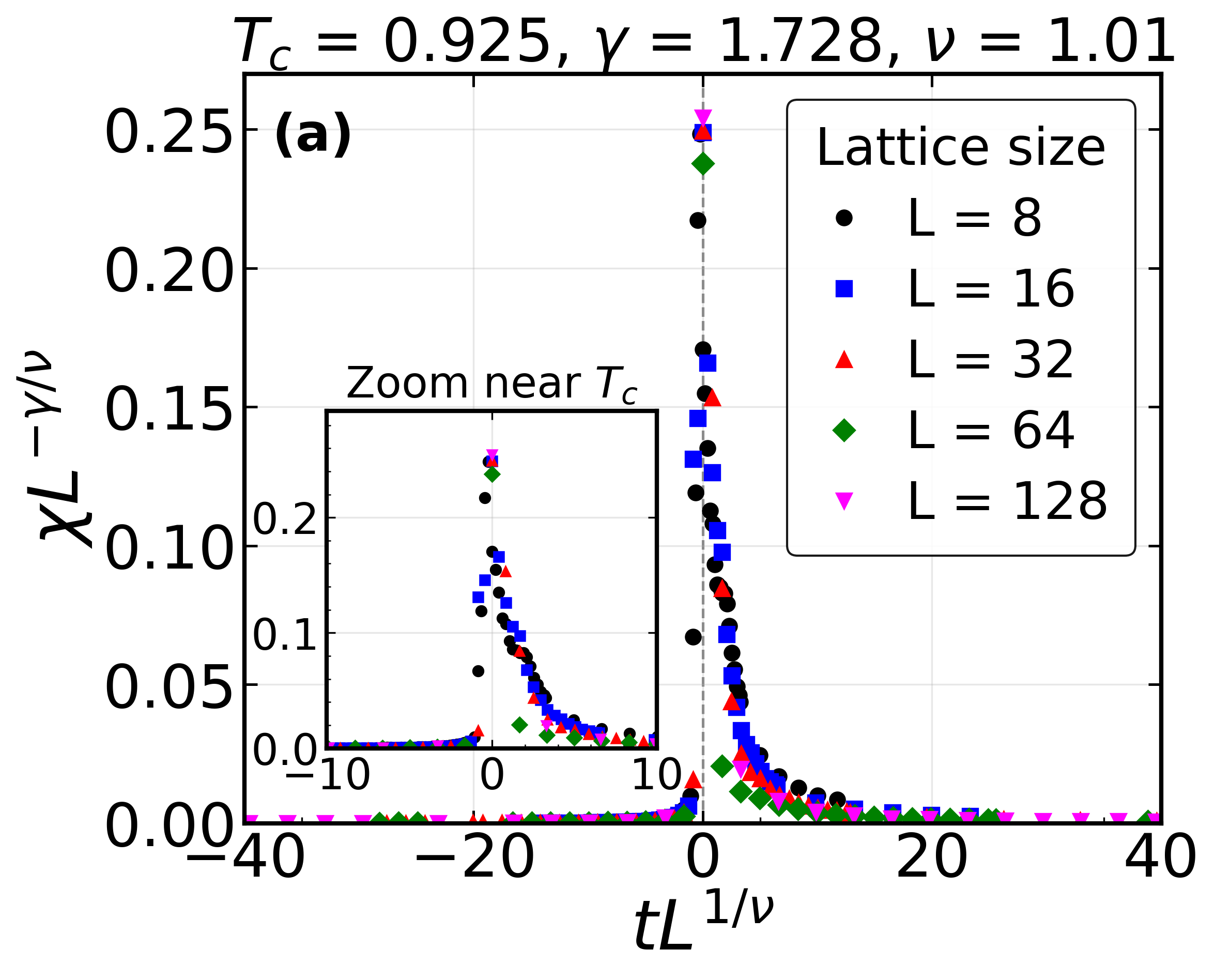}
    \includegraphics[width=0.49\linewidth]{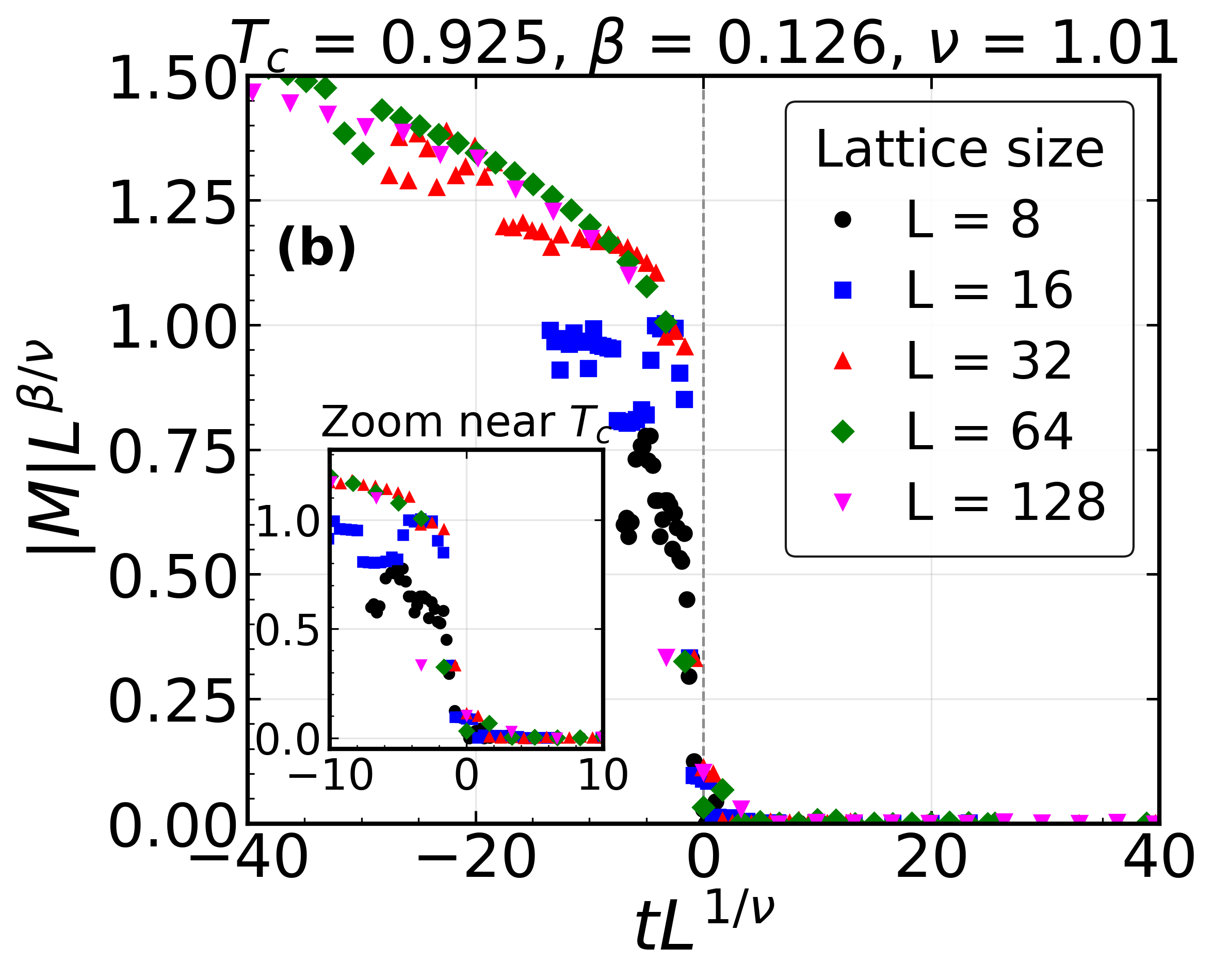}
    \includegraphics[width=0.49\linewidth]{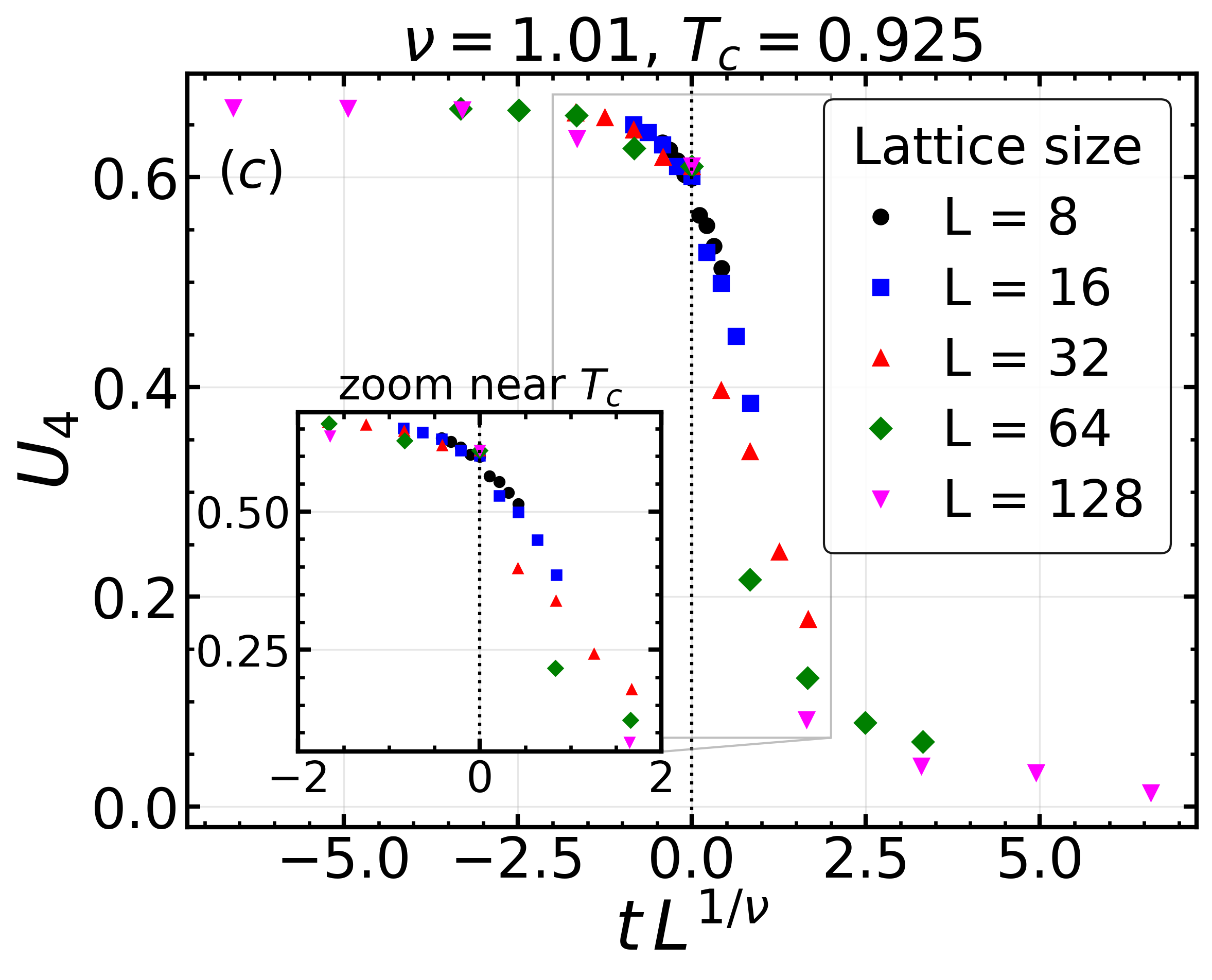}
    \caption{Scaling-collapse plots for $(a)$ Susceptibility, $(b)$ Magnetization and 
    $(c)$ fourth-order Binder cumulant with $T_c=0.925$; $\nu=1.01$, $\gamma = 1.728$, $\beta = 0.126$.}
    \label{fig:actual-scaling-collapse}
\end{figure}

We have also constructed the scaling-collapse plots for $\chi$, $M$ and $U_4$ with our estimated values for the critical exponents, namely, $\gamma = 1.728 $, $\beta = 0.126$ and $\nu = 1.01$ (from Binder cumulant slope) 
as in Fig. \ref{fig:actual-scaling-collapse} (a), (b), (c). This behaviour provides further evidence for the consistency of our results with the two-dimensional Ising universality class.

\section{Summary and perspective}
In this work, we have studied the finite temperature phase transition in the context of $2d$ continuous
spin Ising model with bounded spin $s_i \in [-1,+1]$, 
no external field and no explicit on-site potential.
One of the important aspect of our present study is that an explicit on-site potential is not necessary to generate a finite-temperature phase transition in a bounded continuous-spin system.
The bounded domain of spin can be interpreted as an effective
confining potential giving rise to
critical like scaling of magnetization and susceptibility and the
emergence of Ising universality.
The bounded domain basically acts as an effective constraint on
amplitude fluctuations and it plays a role similar to that of a stabilizing local potential in conventional soft-spin models with a finite-temperature phase transition.

Using Monte Carlo simulations and finite-size scaling
analysis, we have demonstrated the existence of a well-defined second-order phase transition at a critical temperature $T_c \approx 0.925$ in the above bounded continuous spin $2d$ Ising model. 
As an independent check, we have also performed the Binder cumulant analysis to estimate the critical temperature in our model and we find that $T_c \approx 0.925$, which agrees with the value obtained from the analysis of the peak values in the thermodynamic variables. We have also estimated the value of the correlation function critical exponent $\nu$ from the Binder cumulant data.
From the finite-size scaling analysis, we have obtained the critical exponents which are consistent with the values for the
two-dimensional Ising universality class. 
We have also investigated  the finite-size scaling behaviour of magnetization, susceptibility and fourth-order Binder cumulant using the corresponding scaling-collapse plots. This confirms that the model belongs to the same universality class despite the continuous nature of the spins and no explicit on-site potential.
The present model provides a minimal framework which interpolates between the Gaussian model with unbounded spins
having no finite-temperature phase transition, and the discrete Ising model with fixed spin amplitude and a well defined phase transition. The observed critical temperature $T_c$ can be understood in terms of amplitude fluctuations, which reduce the effective interaction strength compared to
the discrete Ising model while still allowing a finite-temperature phase transition.

There are several interesting directions in which the present model can be
explored further. In particular,
one can introduce an explicit on-site potential, such as a quartic potential as discussed in the introduction
and systematically study how the critical
temperature $T_c$ evolves with the strength and shape of the local potential. As another extension of our work, we can consider 
other boundary conditions apart from the periodic one to study the 
phase transition and critical behaviour and compare with the present formulation. 
One may also explore the effect of external field and anisotropic
couplings to understand the continuous spin models further. Another important direction
is to study the above continuous bounded spin model in the framework of
Cellular Automata which has been discussed earlier for standard $2d$
Ising model by two of the present authors
\cite{JM-SM:2018}. 
We hope that the present results will motivate further studies of
bounded continuous-spin systems for exploring the critical phenomena.
 \vspace{-0.5cm}

\section*{Acknowledgment}
\noindent
RM would like to acknowledge Odisha State Higher Education Council, Government of Odisha for the support under Mukhyamantri Research and Innovation
(MRI)-2024 (24EM/PH/102). We thank the anonymous referees for the valuable comments and suggestions. 
 \vspace{-0.5cm}







\bibliography{reference.bib}

\end{document}